\documentclass[aps,prb,showpacs,
floatfix,
amsmath,amssymb,
               superscriptaddress,
               reprint]{revtex4-1}
\usepackage{bbold}
\usepackage{color}
\usepackage{graphicx}
\usepackage{natbib}
\usepackage{setspace}
\usepackage{amsmath}
\usepackage{amssymb}
\usepackage{verbatim}
\usepackage{bibentry}

\definecolor{scarred}{rgb}{0.75,0.0,0.0}

\begin{document}


\title{Unraveling the role of V-V dimer on the vibrational properties of VO$_2$ by first-principles simulations and Raman spectroscopic analysis}
\author{Wasim Raja Mondal}\email{wmondal@binghamton.edu}
\affiliation{Department of Physics, Applied Physics, and Astronomy, Binghamton University,State University of New York, Binghamton, New York 13850, USA.}
\author{Egor Evlyukhin}
\affiliation{Department of Physics, Applied Physics, and Astronomy, Binghamton University,State University of New York, Binghamton, New York 13850, USA.}
\author{Sebastian A. Howard}
\affiliation{Department of Physics, Applied Physics, and Astronomy, Binghamton University,State University of New York, Binghamton, New York 13850, USA.}
\author{Galo J. Paez}
\affiliation{Department of Physics, Applied Physics, and Astronomy, Binghamton University,State University of New York, Binghamton, New York 13850, USA.}
\author{Hanjong Paik}
\affiliation{Departments of Materials Science and Engineering, Cornell University, Ithaca, New York 14853-1501, USA} 
\affiliation{Platform for the Accelerated Realization, Analysis and Discovery of Interface Materials (PARADIM), Cornell University, Ithaca, New York 14853, USA}
\author{Darrell G. Schlom}
\affiliation{Departments of Materials Science and Engineering, Cornell University, Ithaca, New York 14853-1501, USA}
\affiliation{Kavli Institute at Cornell for Nanoscale Science, Ithaca, New York 14853, USA}
\author{Louis F J Piper}
\affiliation{Department of Physics, Applied Physics, and Astronomy, Binghamton University,State University of New York, Binghamton, New York 13850, USA.}
\author{Wei-Cheng Lee} \email{wlee@binghamton.edu}
\affiliation{Department of Physics, Applied Physics, and Astronomy, Binghamton University,State University of New York, Binghamton, New York 13850, USA.}
\begin{abstract}
We investigate the vibrational properties of VO2, particularly the low temperature M1 phase by first-principles calculations using the density functional theory as well as Raman spectroscopy. We perform the structural optimization using SCAN meta-GGA functional and obtain the optimized crystal structures for metallic rutile and insulating M1 phases satisfying all expected features of the experimentally derived structures. Based on the harmonic approximation around the optimized structures at zero temperature, we calculate the phonon properties and compare our results with experiments. We show that our calculated phonon density of states is in excellent agreement with the previous neutron scattering experiment. Moreover, we reproduce the phonon softening in the rutile phase as well as the phonon stiffening in the M1 phase. By comparing with the Raman experiments, we find that the Raman-active vibration modes of the M1 phase is strongly correlated with the V-V dimer distance of the crystal structure. Our combined theoretical and experimental framework demonstrates that Raman spectroscopy could serve as a reliable way to detect the subtle change of V-V dimer in the strained VO$_2$.
\end{abstract}
\maketitle





\section{Introduction}
Vanadium oxide (VO$_2$) has been a focus of intense research since its discovery in 1959\cite{PhysRevLett.3.34}. Besides academic interest, it underpins a plethora of applications including gas sensors\cite{doi:10.1021/nl900676n}, window coatings\cite{doi:10.1021/cm034905y}, resistive random access memory (RRAM) devices \cite{KIM201333}, electronic switches\cite{doi:10.1063/1.324047,doi:10.1063/1.4757865,doi:10.1146/annurev-matsci-062910-100347}, and so on. The versatile phase diagram of VO$_2$ showcase well known first order metal-insulator transition (MIT) at almost room temperature (340 K) and ambient pressure accompanied by a structural transition from a high-temperature rutile (R phase) to a low-temperature monoclinic (M$_1$ phase). The intricate concomitant nature of these two transitions has fueled a long-standing debate on the plausible mechanism for the MIT in VO$_2$. Several scenarios have been adopted including Peierls type\cite{PhysRevB.70.161102,PhysRevLett.72.3389,GOODENOUGH1971490}, Mott-Hubbard-type and even combination of these two\cite{PhysRevLett.94.026404,PhysRevLett.95.196404,PhysRevLett.108.256402}utilizing the role of lattice instabilities, electron-electron interactions, electron-phonon interactions, and so on. Though extensively studied, a consensus for a exact mechanism of MIT in VO$_2$ is still elusive.    

The role of lattice vibrations in this MIT is still not completely understood. Studying lattice vibrational properties of the VO$_2$ turns out to be a non-trivial task from both the experimental and theoretical point of view. For example, the bottleneck of single-crystal inelastic neutron scattering (INS) is the incoherent vanadium neutron scattering cross-section. Even, some of the existing Raman spectroscopic studies of the structural phase transition (SPT) in epitaxial VO$_2$ films have been found to be inconsistent due to the large thicknesses of studied films (80- 100 nm) or even misinterpreted due to the strong signals from the substrate\cite{Petrov2002,Shibuya2014,Okimura2014,Yang2016}. 

Besides these experimental challenges, achieving an exact theoretical description of the lattice dynamics of VO$_2$ has been a tougher nut to crack. Since lattice vibrational properties are strongly related with the structure of the material, first-principle based methods that take into account the much needed material specific information can be an ideal tool for this task. However, a reliable exchange functional which can correctly describe the vibrational properties of all the phases of VO$_2$ is not readily available and still an going topic of research. It has been found that the standard functional such as local density approximation (LDA) and generalized gradient approximations (GGA) can not describe the structural, electronic and magnetic features of the R and M$_1$ phases simultaneously \cite{PhysRevLett.72.3389,PhysRevB.86.075149}. To ameliorate the inaccuracy of the band structure calculations based on the standard functional, several attempts have been already made to go beyond the LDA or GGA functional. For example, the modified Becke-Johnson (mBJ) exchange potential has been demonstrated to be very efficient for describing the MIT in VO$_2$\cite{PhysRevB.86.075149}. Unfortunately, the Hellmann–Feynman force calculations are not possible using the mBJ ruling out the possibility of phonon calculations within this exchange approximation. The GW method and its variants successfully predict the electronic feature of the MIT\cite{PhysRevB.60.15699,PhysRevLett.99.266402,PhysRevB.78.075106}. However, equivalent GW calculations for phonons are still not clear. The dynamical mean-field theory and its cluster extensions correctly describe the electronic properties of both the R and M$_1$ phases of VO$_2$\cite{PhysRevB.73.195120,PhysRevLett.94.026404}. Cluster extension of DMFT such as dynamical cluster approximations (DCA) and its typical medium extension named as typical medium DCA (TMDCA) for disordered phononic systems has been recently developed\cite{PhysRevB.96.014203,PhysRevB.99.134203}. Unfortunately, such DCA and TMDCA method for phononic systems are currently available for parameter-based model calculations. Moreover, hybrid functionals efficiently describe the electronic properties of  VO$_2$\cite{PhysRevLett.107.016401} and is found to be also successful in describing the structural properties of all the phases of VO$_2$\cite{PhysRevB.95.125105}. However, phonon calculations using hybrid functionals turn out to be  a very expensive computational task. First-principle based Quantum Monte Carlo based calculations have been also performed for VO$_2$\cite{PhysRevLett.114.176401}. Again, computational expenses discard such possibility for applying the method in the lattice dynamics calculations. 

Considering all these limitations, the GGA+U method \cite{LDA+U,PhysRevB.87.195106,PhysRevB.99.064113,Budai2014} has been extensively used for phonon calculations of VO$_2$. However, the GGA+U method remains controversial in producing an accurate description of VO$_2$. Anisimov and co-authors\cite{LDA+U} find insulating nature of the R phase which strongly disagree with experimental findings. Mellan et al.\cite{PhysRevB.99.064113} predict electronic structure of VO$_2$ that is consistent with experiment using non-spin polarised GGA+U calculations. But, U value within the LDA+U method has been treated so far as a free parameter and the results are also very sensitive with the choice of U value. For example, Budai et al.\cite{Budai2014} point out that the U value less than 3.4 eV is more appropriate in determining energy difference between the R and M$_1$ phase that could be more consistent with experiment, but such lower U value can produce more smaller bandgap. Hence, an accurate prediction of the U value rather using it as a free parameter or bare atomic value, may have a paramount importance for reliable estimation of the vibrational properties of VO$_2$. To best of our knowledge, calculations for determining U value of VO$_2$ via the constrained random phase approximation (cRPA) or linear response method are still missing. Another important thing is to include temperature effect and anharmonicity. Such calculations\cite{Budai2014} have been already performed and can help to validate further theoretical development.

In this paper, we present a comprehensive study of the lattice dynamics and electronic structures using various DFT functionals and benchmark our calculations with experiments. 
By comparing with Raman spectroscopy, we investigate the interplay between strain effect and local environment (e.g. spectral shifts of the V-V and V-O phonon modes) in epitaxial VO$_2$ thin films deposited on MgF$_2$ (001) and (110) substrates. We show that our calculations are in a good agreement with the observed Raman shifts of the characteristic phonon modes induced by the epitaxial strains imposed on VO$_2$ thin films. Our work demonstrates that Raman spectroscopy could serve as a reliable way to detect the subtle change of V-V dimer in the strained VO$_2$. 
 
We have organized this paper as follows: we give details of our first-principles simulations in section II. We provide information of our experimental set up in section III. We present and discuss our results in section IV. Finally , we conclude our discussions in section V.

\begin{figure}[h!]

\centerline{\includegraphics[scale=0.4]{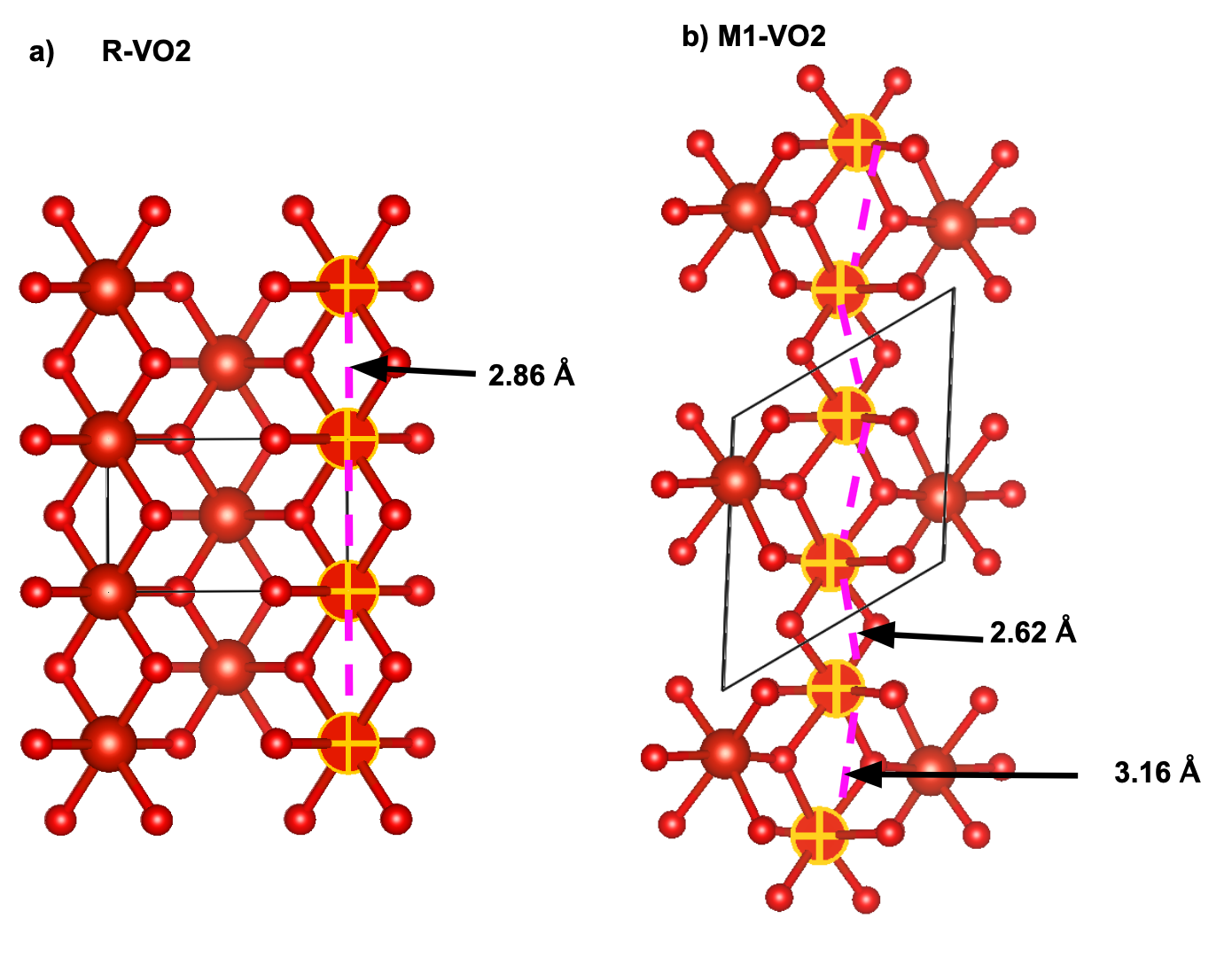}}
\caption{The schematic of the experimental structures\cite{eyrotstructure} for (a) rutile (R), (b) monoclinic M1 phases of VO2. Big (red) and small (red) balls denote vanadium and oxygen atoms, respectively.}
\label{fig:structure}
\end{figure}

\section{Computational details}
We consider a plane-wave basis and the projector-augmented- wave (PAW) method as implemented in the Vienna Ab initio Simulation Package (VASP)\cite{PhysRevB.54.11169,KRESSE199615}. We have used several different exchange correlation functional including the LDA, the PBE GGA\cite{PhysRevLett.77.3865}, and the SCAN\cite{PhysRevLett.115.036402}. We note that the SCAN has been implemented for self-consistent calculations in the VASP and such self-consistent implementation is still not available in other first-principles packages like WIEN2K. We choose PBE GGA pseudopotential  in our calculations using SCAN. For  electronic structure calculations, we use the plane-wave cutoff energy as 500 eV. We also verify that there is no appreciable changes in the results using a cutoff energy of 600 eV. For all Brillouin-zone sampling, $\Gamma$-centered k-point grids is considered as $16 \times 16 \times 16$. The self-consistent calculations are considered to be converged by considering $10^{-4}$ eV between successive iterations, and the convergence of structural relaxation is decided by the total energy difference between two successive ionic steps as $10^{-3}$ eV. We have also cross-checked our electronic structure calculations by using mBJ calculations as available in WIEN2K. The results obtained from the WIEN2K are not shown here. For the phonon calculations, we have employed the pseudopotential band method and the supercell approach that are implemented in VASP\cite{PhysRevB.54.11169,KRESSE199615} and PHONOPY\cite{phonopy}. As a standard supercell based phonon calculations, force-constants are computed by means of the Hellmann-Feynman theorem\cite{PhysRevLett.78.4063}. 

\section{Experimental details}
High quality 10-nm-thick VO$_2$ films were grown on rutile MgF$_2$ (001) and (110) single crystal substrates by reactive molecular-beam epitaxy (MBE) via codeposition method under a distilled ozone background pressure at the PARADIM Thin Film Growth Facility at Cornell university.\cite{Paik2015} Raman microscopy analysis was conducted at the Center for Nanoscale Materials at the Argonne National Laboratory. The spectra were recorded using a Raman microscope (inVia Reflex, Renishaw, Inc.) with spectral resolution of 0.5 cm$ ^{-1} $ using 532 nm excitation from a diode pumped solid state laser (RL532C50, Renishaw, Inc.). Samples were held in a nitrogen-purged temperature-controlled stage (THMS600/TMS94/LNP94, Linkam Scientific Instruments Ltd). Excitation and collection of scattered light occurred through a 50x objective (Leica, NA=0.50). The laser power was set to 0.64 mW to exclude local heating effect. Collected spectra are typically consisted of the average of 20 integrations, where each integration was collected for 1 min.

\section{Results and discussions}

We begin our discussions by understanding the well-known experimental crystal structure of VO$_2$\cite{eyrotstructure} as described in Fig.~\ref{fig:structure}. The unit cell contains 6 atoms for the R phase and 12 atoms for the M$_1$ phase in our considered experimental structure. As shown in Fig.~\ref{fig:structure}(a), the V-V atoms form a periodic V chain with fixed V-V bond distance as 2.86 $\AA$ in the rutile structure, whereas there are significance differences in the arrangement of the V atoms along the rutile $C_R$ axis in the M$_1$ phase, as demonstrated in Fig.~\ref{fig:structure}(b). In this M1 phase, the V atoms are arranged such a way that they form dimmer alternatively and tilt along the $C_R$ axis, which leads to doubling of the unit cell volume of that for the R phase with V-V distance as 2.62 $\AA$(bonding) and 3.16 $\AA$ (anti-bonding). With this understanding of available experimental crystal structure, we perform structure optimization of both the R and M1 phases of VO$_2$.

\begin{table}
\begin{tabular}{l*{6}{c}r}
Functional              & V-V distance  \\
\hline
Exp.\cite{eyrotstructure}    & 2.86   \\
GGA     & 2.76  \\
Ref\cite{PhysRevB.95.125105}.  & 2.80 \\
SCAN    & 2.77  \\
\end{tabular}
\caption{Optimized V-V distance for the rutile phase of VO$_2$ obtained from our calculations using different functional.}
\label{tab:Roptimization}
\end{table}

\begin{table}
\begin{tabular}{l*{6}{c}r}
Functional              & V-V dimer (long) & V-V dimer (short)  \\
\hline
Exp.\cite{eyrotstructure}    & 3.16  & 2.62 \\
LDA     & 2.72 & 2.72 \\
GGA     & 3.0 & 2.59 \\
Ref\cite{PhysRevB.95.125105}.  & 3.14 & 2.44 \\
SCAN    & 3.16 & 2.47 \\
\end{tabular}
\caption{\label{tab:M1optimization}Optimized long V-V dimmer and short V-V dimer for the M$_1$ phase of VO$_2$ obtained from our calculations using different functional.}
\end{table}

\begin{figure}[h!]
\centerline{\includegraphics[scale=1.2]{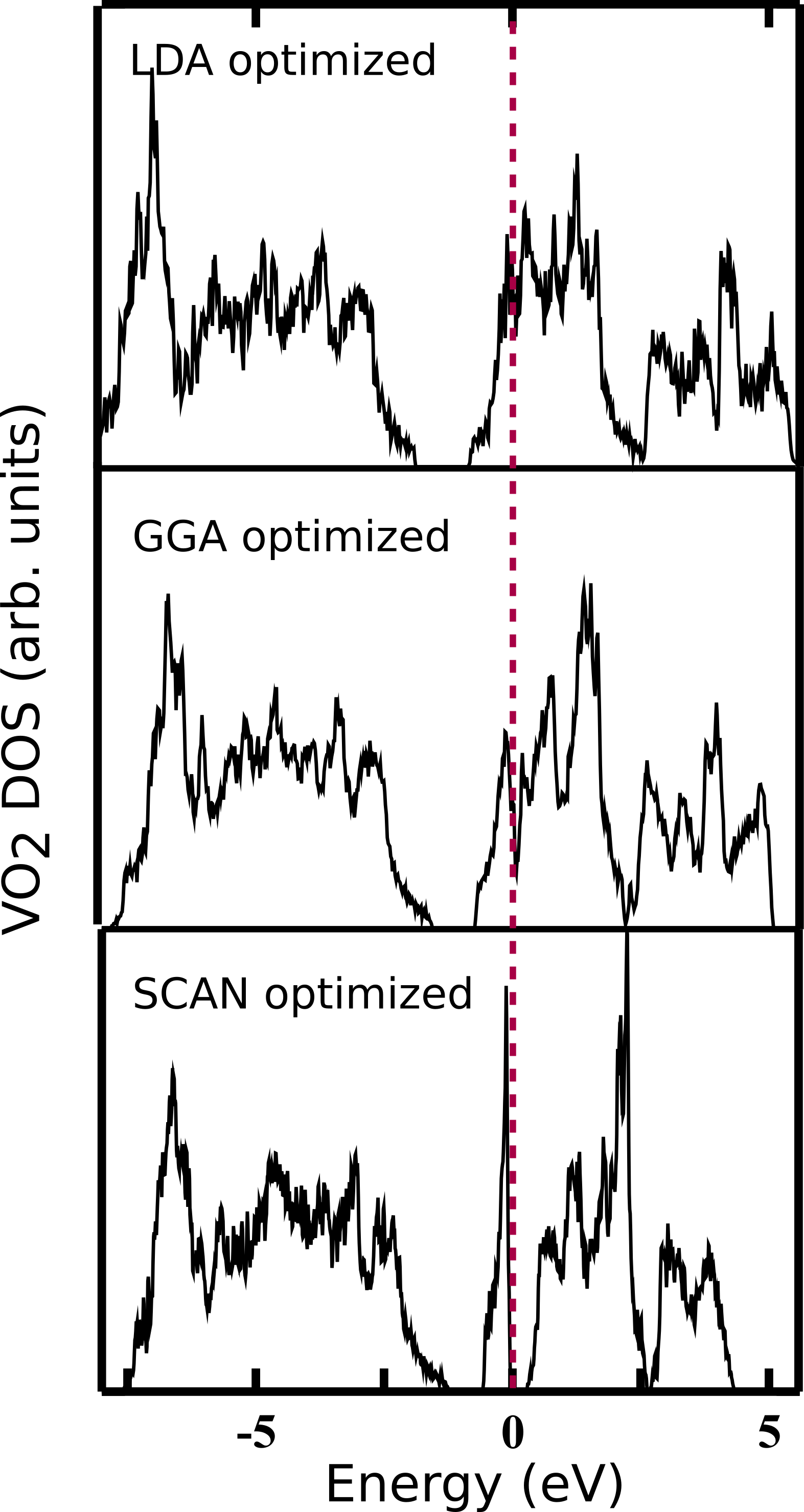}}
\caption{Comparison of the electronic density of states (DOS) from the LDA, GGA and SCAN optimized structure (from top to bottom of the figure). Considering optimized structure with different functional as input, all DOS calculations are performed by using SCAN. The gapped DOS with finite band gap value is only obtained from the SCAN optimized structure.}
\label{fig:dosfromopt}
\end{figure}

\begin{figure}[h!]
\centerline{\includegraphics[scale=1.2]{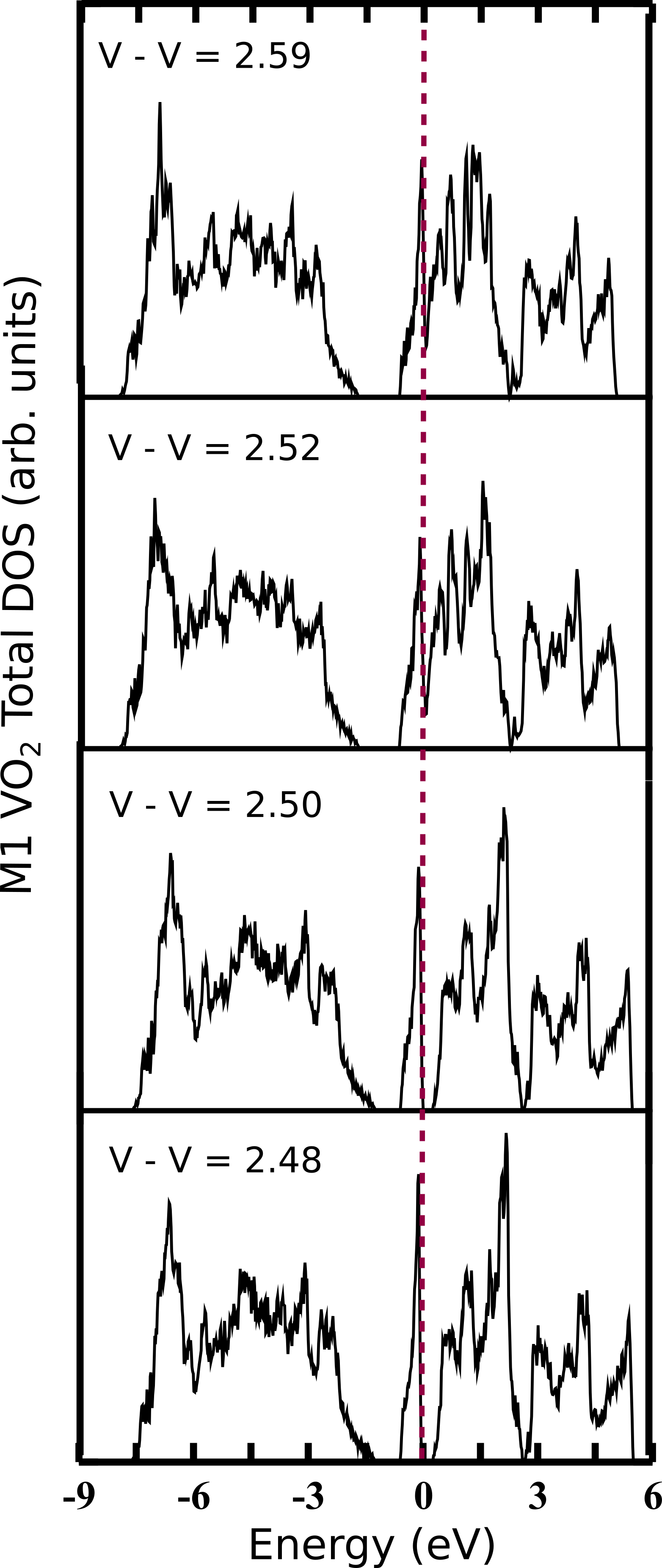}}
\caption{The evolution of electronic density of states (DOS) with variation of short V-V dimer distance.In all four calculations of DOS, SCAN meta-GGA functional has been used. A gap opens at a critical value of the short V-V dimer distance suggesting that structure property of the M$_1$ phase play a significant role in controlling its insulating nature.}
\label{fig:dosbandinsultor}
\end{figure}

\begin{figure}[h!]
\centerline{\includegraphics[scale=0.8]{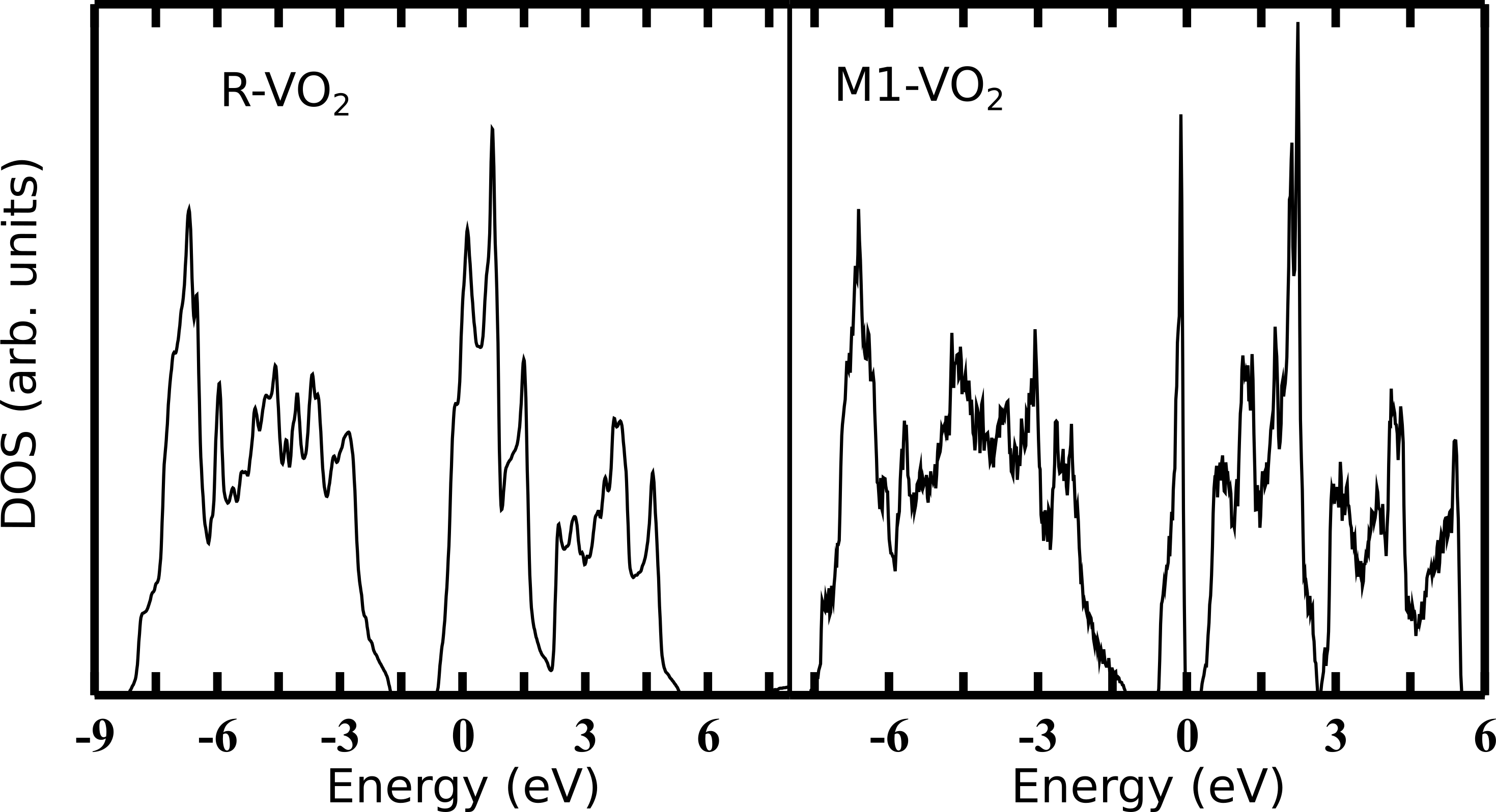}}
\caption{Electronic density of states (DOS) for high-temperature rutile (R) and low temperature monoclinic (M1) phases of VO$_2$ obtained from using SCAN meta-GGA functional. In calculations for both the phases, we have used the SCAN optimized structure. The SCAN simultaneously predict R phase as a metal and M$_1$ phase an insulator, consistent with previous theories and experiments.}
\label{fig:RM1scandos}
\end{figure}

For structure optimization, we consider the ionic optimization process, in which the lattice parameters are fixed and atoms are allowed to move to search for minimum energy configurations. First, we optimize the crystal structure of the R phase using different functional and compare with experiment\cite{eyrotstructure} as well as previously reported values\cite{PhysRevB.95.125105} as given in  Table~\ref{tab:Roptimization}. For SCAN meta-GGA functional, the V-V pair distance changes from the experimental value 2.86 $\AA$ to 2.76 $\AA$, whereas it is 2.77 $\AA$ for the GGA functional. We note that there is no significant changes in the results obtained from the SCAN meta-GGA and GGA functional. However, this is certainly not true for the case of M$_1$ phase. Our calculations suggest that the optimization process is extremely sensitive to the functional used in the calculations as summarized in Table~\ref{tab:M1optimization}. We find that the LDA functional completely fails to optimize the M1 structure. We observe that the long V-V dimer and short V-V dimer distance becomes almost equal as 2.72 $\AA$ which imply that the M1 structure reduces to the rutile like structure through the LDA optimization process. Hence, like electronic properties, our calculations reveal that the LDA is not applicable to determine the structural properties of the M1 phase. The situation becomes better when the GGA is applied in our calculations. In the optimized M1 structure based on the GGA functional, the long V-V dimer distance decreases to 3.0 $\AA$ from the experimental value $3.16 \AA$ and short V-V dimer separation decreases to $2.59 \AA$ from the experimental value $2.62 \AA$. Such drastic change in the dimmer distance clearly indicates that the M1 structure again tends to go to the rutile phase through the optimization process. As has been already found\cite{PhysRevB.95.125105}, this failure of the LDA and GGA can be cured by the Hybrid functional. To overcome this barrier, we have utilized SCAN meta-GGA functional for optimizing the M$_1$ structure. As summarized in  Table~\ref{tab:M1optimization}, in the SCAN optimized M1 structure, the long V-V dimer nature survive as $3.16 \AA$ which is very close to the value $3.14$ as found in hybrid calculations\cite{PhysRevB.95.125105} and exact as $3.16 \AA$ as found in experiment\cite{eyrotstructure}. The short V-V dimer using SCAN functional is found to be $2.47 \AA$ which is also to $2.44 \AA$ as found from the hybrid functional calculations. Thus, like hybrid functional, the optimized crystal structure using SCAN has all the expected features of the experimentally derived structure: All V-V chains of M1 are found to be dimerized and canted along the rutile axis, whereas all V-V chains form straight undimerized one in the rutile phase.

Here we want to investigate the effect of the structural optimization on the electronic structure, in order to draw an accurate band theory picture of the M1 phase. Fig. 2 presents the electronic density of states (DOS) based on our optimized structures. To treat the comparison on equal footing, we use SCAN functional in the calculation of electronic DOS from the structures that are obtained from different functional. As shown in the top panel of Fig.~\ref{fig:dosfromopt}, there is no gap in the DOS corresponding to the LDA optimized structure. However, a pseudogap feature tends to emerge in the DOS of the GGA optimized structure as shown in the middle panel of the Fig.~\ref{fig:dosfromopt}. Such small gap in the DOS with the GGA optimized structure was also previously reported and our calculations are consistent with these previous calculations\cite{PhysRevB.87.195106,Budai2014}. Interestingly, we find a fully gapped spectrum with the SCAN optimized structure as displayed in the bottom panel of Fig.~\ref{fig:dosfromopt}. 

\begin{figure}[h!]
\centerline{\includegraphics[scale=0.7]{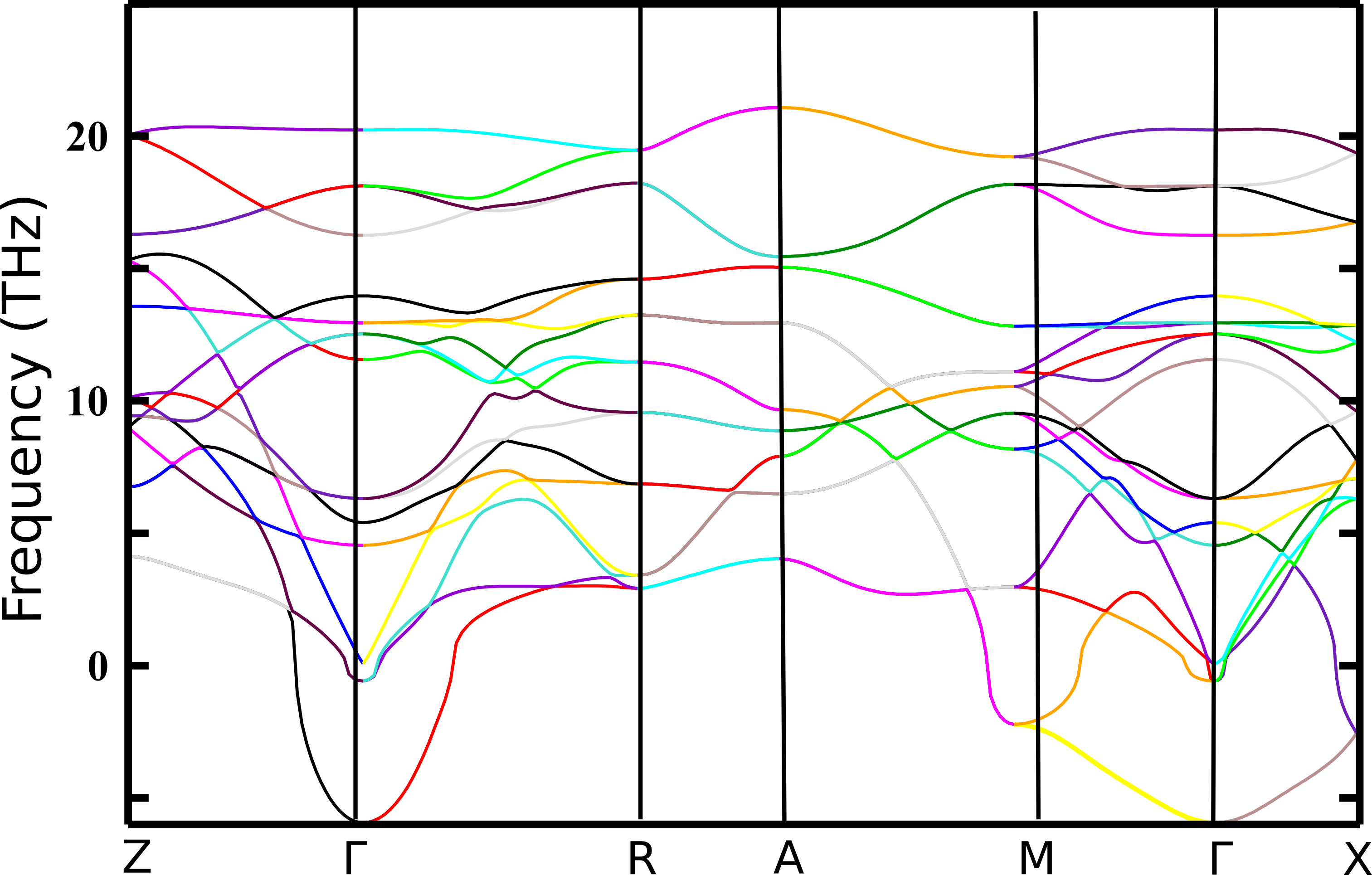}}
\caption{Phonon dispersions of the R-VO$_2$ calculated using the SCAN meta-GGA functional within the zero temperature and harmonic approximation. The negative phonon frequencies (imaginary frequencies) indicating the phonon softening instability of the R phase at low temperature.}
\label{fig:rutilband}
\end{figure}

\begin{figure}[h!]
\centerline{\includegraphics[scale=0.7]{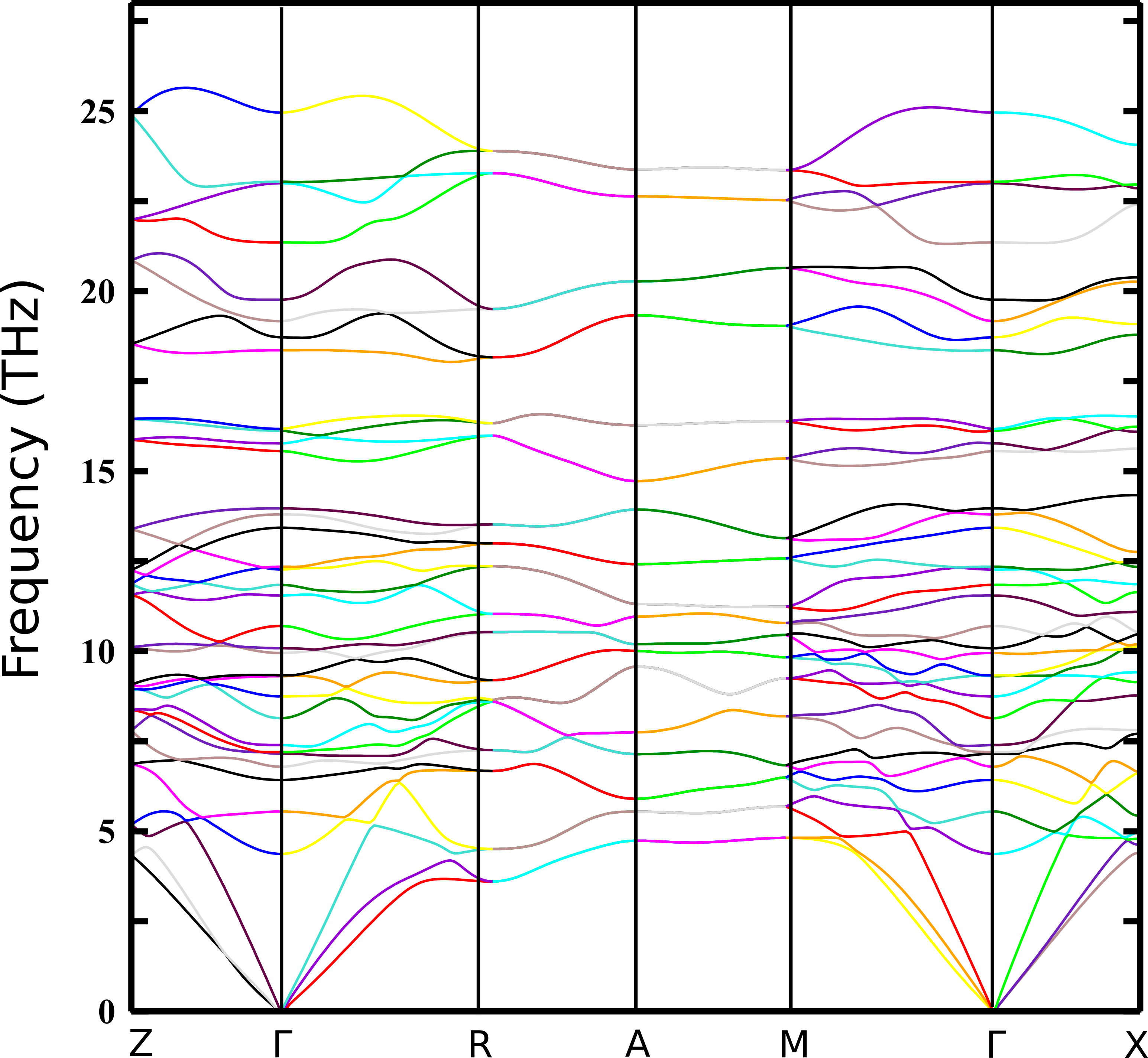}}
\caption{Phonon dispersions of the M$_1$-VO$_2$ calculated using the SCAN meta-GGA functional within the zero temperature and harmonic approximation. The absence of any negative phonon frequencies (imaginary frequencies) suggesting the stability of the M$_1$ structure at low temperature.}
\label{fig:M1band}
\end{figure}

Next, we study the effect of V-V dimerization on the electronic properties of M$_1$ phase. Fig.~\ref{fig:dosbandinsultor} displays the variation of the density of states as a function of various values of short V-V dimer separation. As it is clearly seen from the top to bottom of this figure, there is no gap in the density of states for the V-V dimer distance as $2.59 \AA$. Interestingly, a finite gap emerges as the V-V dimer separation is varied from $2.59 \AA$ to $2.48 \AA$. Thus, our calculations suggest that a gap can be opened up for the M$_1$ phase depending only on the structure property without considering any Hubbard on-site  coulomb interactions. From our calculations, it is certainly clear that the role of Peierl's distortion cannot be neglected in determining the insulating nature of the M$_1$ phase. Such observations strongly suggest that the M$_1$ is not pure a conventional Mott insulator.



Besides successfully describing $M_1$ phase, determining the correct electronic structure of VO$_2$ including both rutile and M$_1$ phases using a single-functional is a considerable difficult task from band-theory point of view. To test the  performance of SCAN against other functional in describing both the rutile and M1 phases of VO$_2$, we calculate the density of states as presented in Fig.~\ref{fig:RM1scandos}. From the optimized structure, SCAN based calculations predict the high temperature rutile phase as a metal and the low temperature M1 phase as an insulator. These findings are consistent with experiments. We note that previous calculations based different functional fail to correctly describe both the phases of VO$_2$ simultaneously. For example, the LDA+U method find insulating nature of the rutile phase of VO$_2$\cite{LDA+U,Liu2010}. However, mBJ and hybrid functional as we discussed before can successfully describe both VO$_2$ phases. As we have already mentioned, mBJ cannot be utilized for calculating the forces and hybrid functional is computationally very expensive. Hence, they are not highly suitable particularly for phonon calculations of VO$_2$.



After establishing the correctness of our optimized structure and producing consistent electronic properties with the SCAN optimized structure, now we are ready to discuss our phonon results. Fig.~\ref{fig:rutilband} shows the phonon dispersion curves of R-VO$_{2}$ using SCAN meta-GGA functional. We have also performed same calculations using GGA functional. We find that the phonon dispersion for the rutile phase using the SCAN does not deviate significantly from the results using the GGA functional and observe the phonon softening instabilities in both the SCAN and GGA calculations. Such phonon softening of the rutile phase has already been reported before \cite{PhysRevB.87.195106}. We note that zero temperature DFT calculations within harmonic approximations fail to predict stable rutile phase which is also expected since the rutile phase is stable only at high temperature. We obtain phonon softening at $\mathbf{q}=\mathbf{\Gamma}, \mathbf{M}$ and $\mathbf{X}$ which are excellent agreement with previous GGA based calculations\cite{PhysRevB.87.195106}. Existing studies \cite{PhysRevB.1.2557,Hearn_1972,PhysRevB.10.490,PhysRevB.17.2494,PhysRevB.31.4809,WOODLEY2008167} 
suggest that the phonon softening at $\mathbf{q=R}$ leads structural transition from the high temperature R structure to low temperature $M_1$ structure. 
However, we have not observed any phonon softening at $\mathbf{q} = \mathbf{R}, \mathbf{A}$ and $\mathbf{Z}$. The phonon softening at $\mathbf{q}=R$ is observed in the previous LDA+U calculations\cite{PhysRevB.87.195106}. It has been argued electron correlation is the driving force for such instability. However, recent experimental study\cite{Budai2014} shows that experimental dispersion curve with no negative frequencies of the rutile phase can be reproduced by including temperature and anharmonicity in the theoretical calculations. However, the role of structural instability in driving the MIT is still not well understood. 

\begin{figure}[h!]
\centerline{\includegraphics[scale=0.7]{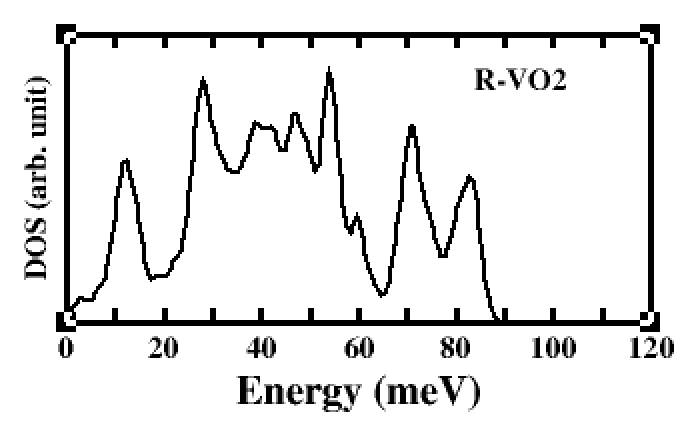}}
\caption{The phonon density of states (PDOS) of the  R-VO$_2$ obtained from zero temperature harmonic apprximation based calculations. Our computed PDOS for the R-VO$_2$ is in agreement with neutron scattering data\cite{Budai2014}.}
\label{fig:natcomrutile}
\end{figure}

To test the limitations of harmonic approximations and zero temperature calculations, we calculate phonon density of states of rutile phase as shown in Fig.~\ref{fig:natcomrutile}. Followed by the dispersion, there are negative frequencies in the low energy region of the spectrum which is not shown here. Interestingly, we obtain a good agreement with the experimental phonon density of states. We note that the experimental phonon spectrum has been collected at temperature T = 381 K, whereas our calculations are limited to T = 0 K within harmonic approximations. Despite these limitations, we are able to capture qualitatively peak features of the experimental spectrum around 18 meV, 33 meV, 45 meV, and 72 meV. We also observe that the height of the peak around 18 meV and 33 meV is more compared to the experimental one. This could be consequence of temperature effect. In addition, we also compare our results against previous calculations\cite{Budai2014}. As we expected, we notice that the height of the peak decreases significantly particularly around 18 meV and 33 meV as temperature increases from 0 to 425 K. We also observe that our calculations are able to capture some features of the spectra that agree better with the experimental spectrum compared to previous theoretical calculations\cite{Budai2014}. For example, the peak around 55 meV corresponding to the experimental DOS is shifted to around 65 meV in the DOS obtained from the previous calculations\cite{Budai2014}. Also, previous calculations find DOS at the high energy region remain around 83 meV, whereas the bandwidth of the experimental spectrum is larger. We note that the bandwidth of our calculated spectrum is much closer to the band width of the experimental DOS compared to previous finite temperature calculations. Although the high temperature rutile phase is believed to be anharmonic, the low temperature M1 phase is found to be highly harmonic. So, we expect our theoretical framework is more suitable for the M1 phase, which is our next topic of discussion. 

\begin{figure}[h!]
\centerline{\includegraphics[scale=0.7]{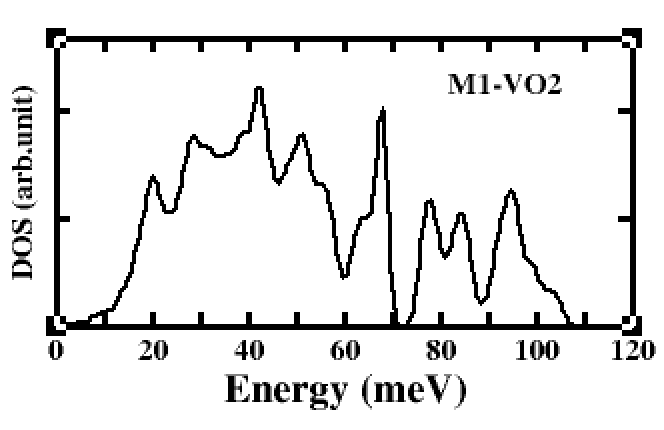}}
\caption{The phonon density of states of the M$_1$-VO$_2$ obtained from zero temperature harmonic approximation based calculations. Our computed PDOS of the M$_1$-VO$_2$ is in excellent agreement with the neutron scattering data \cite{Budai2014}.}
\label{fig:naturecompare}
\end{figure}

\begin{figure}[h!]
\centerline{\includegraphics[scale=0.9]{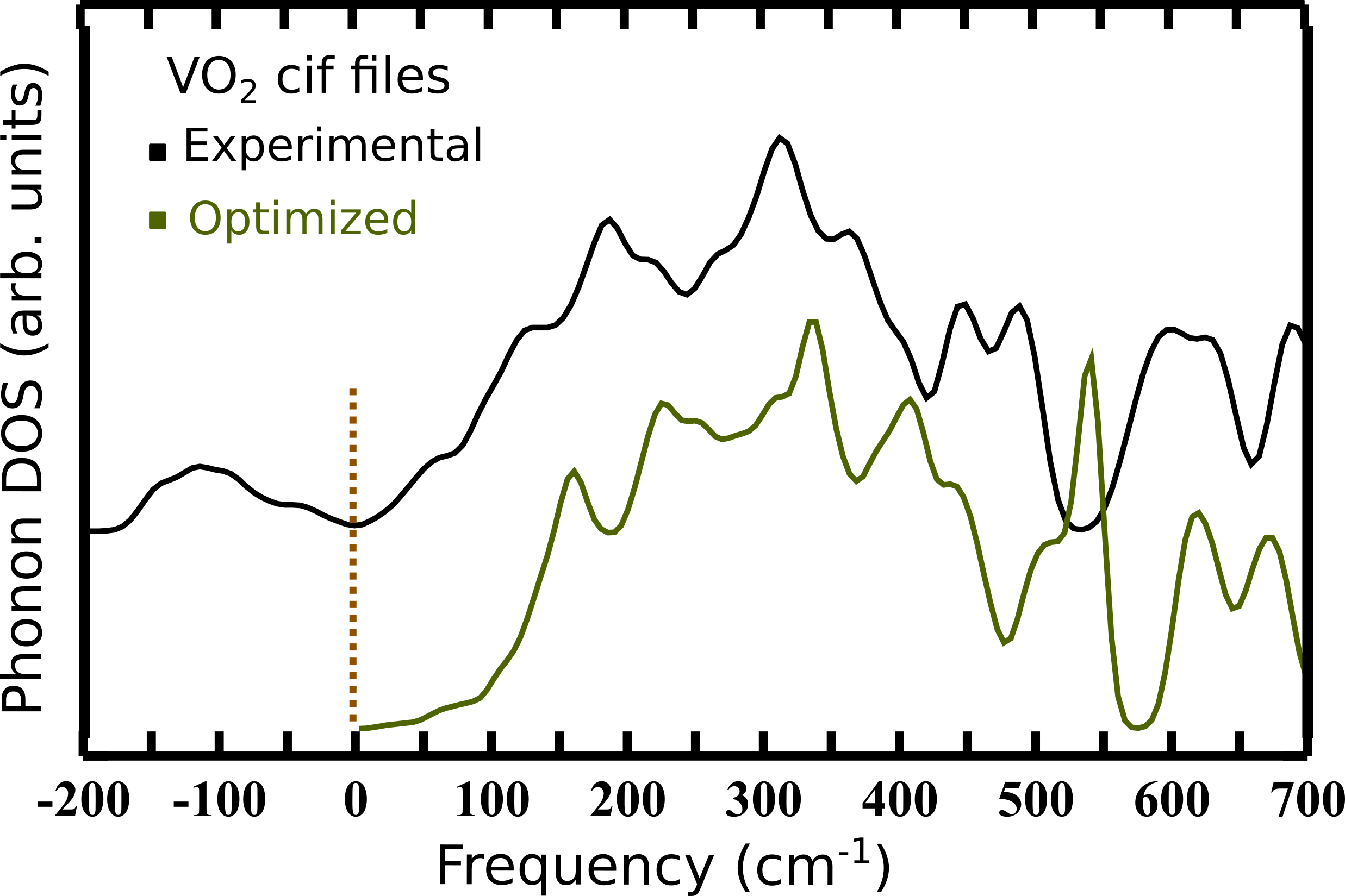}}
\caption{Comparison of the phonon density of states of the M$_1$ phase of VO$_2$ obtained from the experimental structure (top) and the optimized structure(bottom). In both calculations, SCAN meta-GGA functional has been used. In the optimized structure, there is no negative frequency showing stability of the structure, however, the peak position deviates from the positions as found in Raman spectra. But, this agreement can be improved by considering experimental structure as shown in the upper panel. But, there be negative frequency associated with the phonon density of states if phonon calculations are done with out optimizing the experimental structure as expected.}
\label{fig:ramancompare}
\end{figure}

\begin{figure}[h!]
\centerline{\includegraphics[scale=1.1]{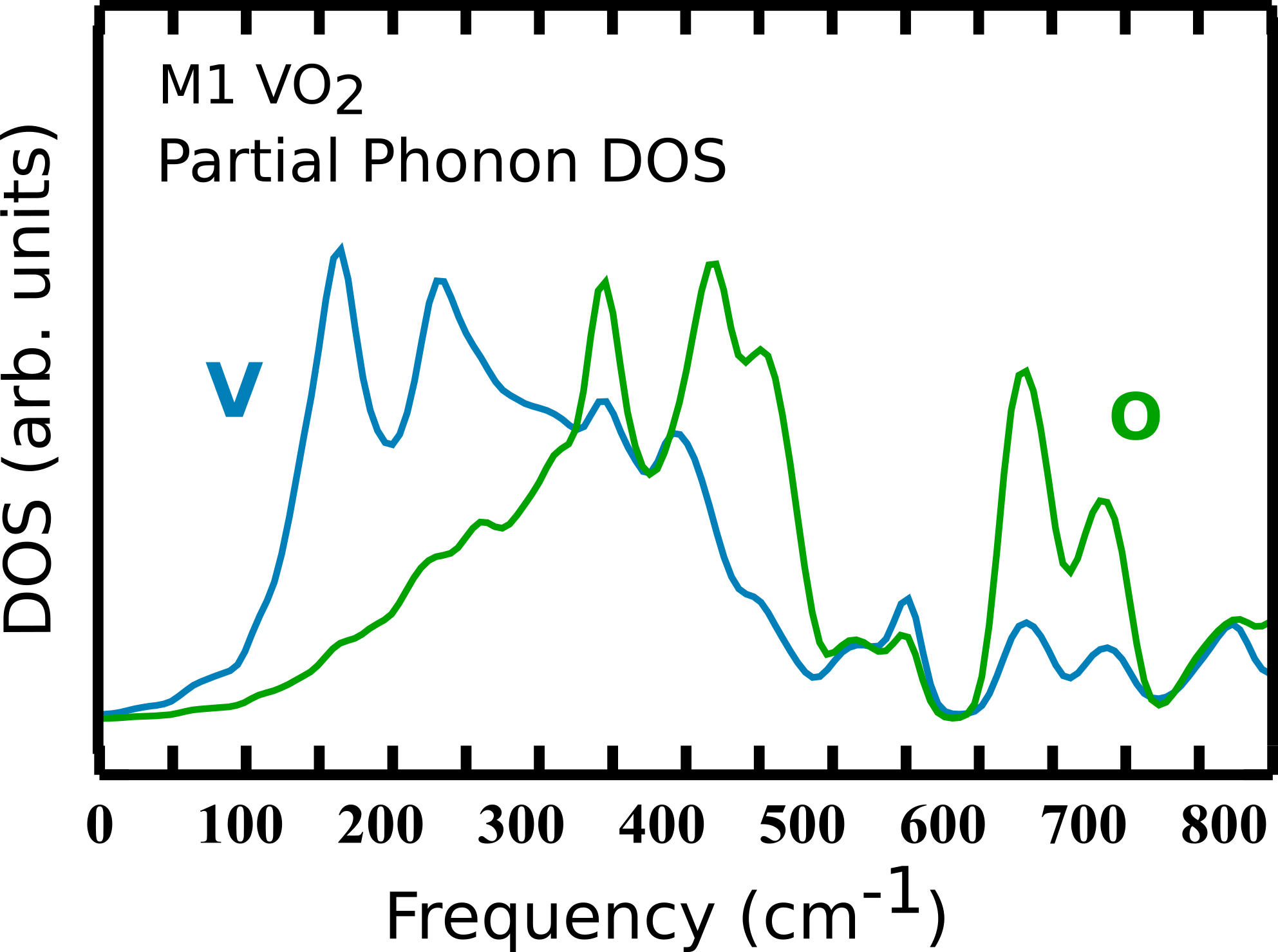}}
\caption{Partial phonon density if states for V and O atoms of M1 phase. The vibrational density of states at the low frequency region around 190 cm$^{-1}$ is mainly dominated by V atoms. However, at the high frequency region around 609 cm$^{-1}$, there is significant contribution from both the V and O atoms.}
\label{fig:partialdosM1}
\end{figure}

\begin{figure}[h!]
\centerline{\includegraphics[scale=0.6]{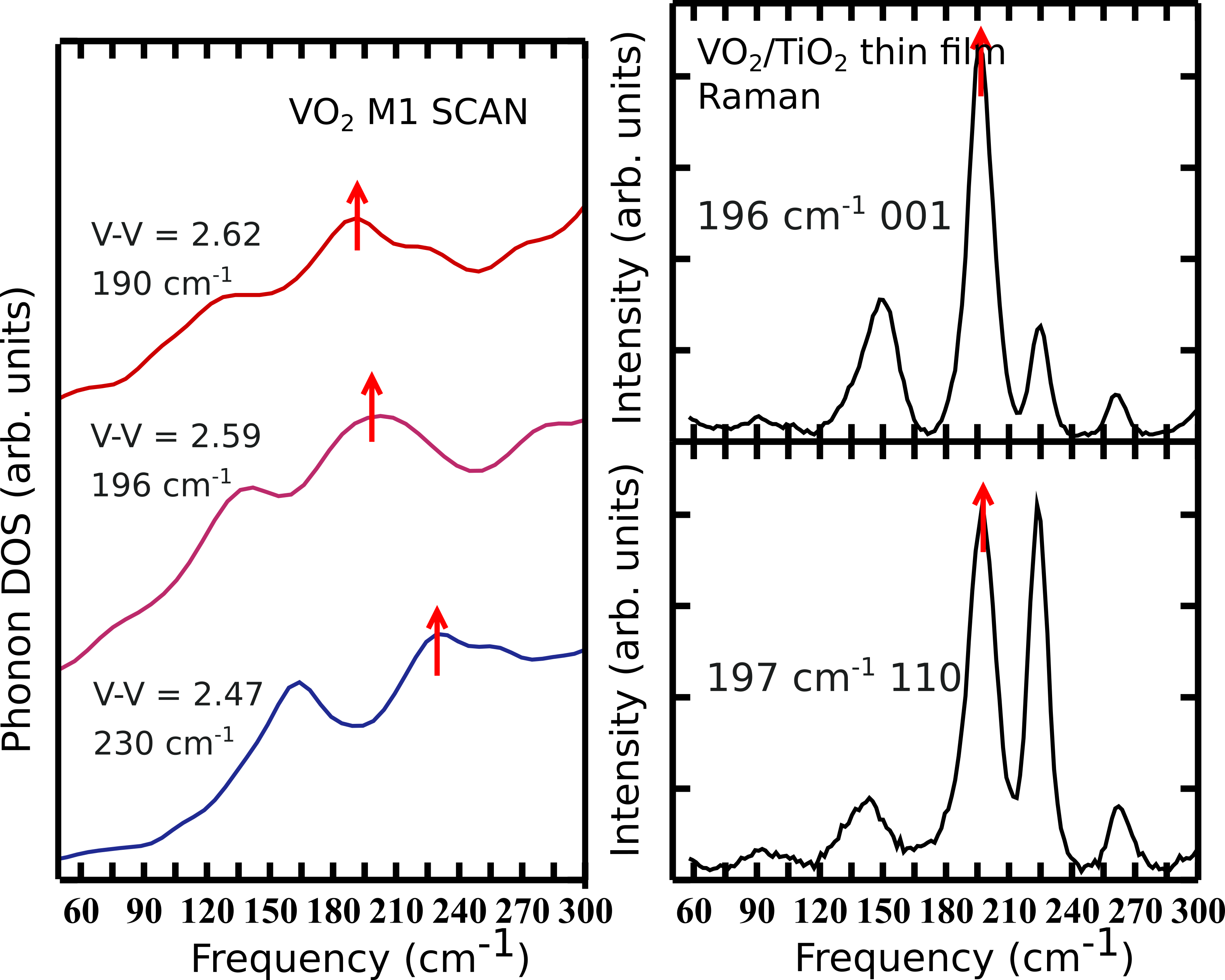}}
\caption{Left panel: Evolution of the low-frequency peak around 190 cm$^{-1}$ in the phonon density of states at different values of the V-V dimer distance of the M$_1$ bulk structure. Right panel: Evolution of the low-frequency peak around 196 cm$^{-1}$ in the Raman spectrum at different values of strain on the 001 and 110 surface of M$_1$ structure. Our calculated value of the peak-position around 190 cm$^{-1}$ is also in strong agreement with previous theoretical and experimental estimated value. First-principles based simulations qualitatively capture the experimental observation of the Raman spectra at the low frequency region with strain engineering.}
\label{fig:lowramancompare}
\end{figure}

\begin{figure}[h!]
\centerline{\includegraphics[scale=0.6]{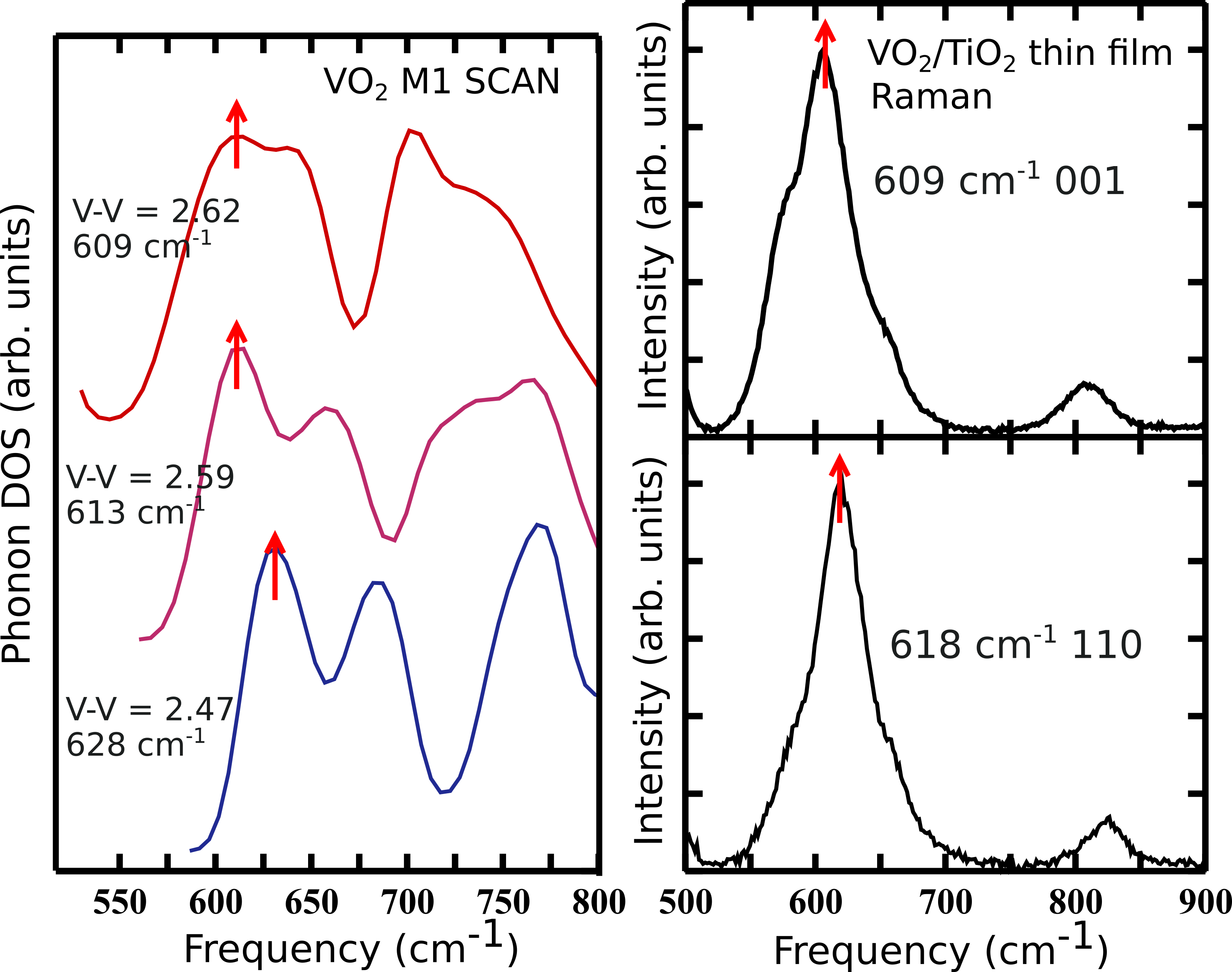}}
\caption{Left panel: Evolution of the high-frequency peak around 609 cm$^{-1}$ in the phonon density of states at different values of the V-V dimer distance of the M$_1$ bulk structure. Right panel: Evolution of the low frequency peak around 609 cm$^{-1}$ in the Raman spectrum at different values of strain on the 001 and 110 surface of M$_1$ structure.
Our computed value of the peak position around 609 cm$^{-1}$ is in strong agreement with previous theory and experimental calculated value. First-principles based simulations qualitatively capture the experimental observation of the Raman spectra at the high frequency region with strain engineering.}
\label{fig:highramancompare}
\end{figure}


We have computed phonon dispersion of the M$_1$ phase as displayed in Fig.~\ref{fig:M1band}. Our calculated dispersion curve for the M1 phase is in good agreement with previous calculations\cite{Lee371}. As shown in Fig.~\ref{fig:M1band}, unlike rutile phase, there is no phonon softening which is in agreement with experimental observations\cite{Budai2014}. It confirms that this M1 phase is the stable structure of VO$_2$ at low temperature. Such structural information can be directly probed in Raman spectra. To understand that, a careful analysis of the phonon density of states can be good a starting point.

The phonon density of states of M$_1$ phase is presented in Fig.~\ref{fig:naturecompare}. Followed by dispersion, there is no negative frequency in the low frequency region of our calculated phonon density of states. We compare our calculated phonon density of states at T =0 K with the phonon density states corresponding to T=298 K obtained from experiment using neutron spectrometer as reported in Ref\cite{Budai2014}. We find our zero temperature calculations using SCAN functional successfully capture the experimentally observed phonon density of states at room temperature. To check the performance of SCAN against other functional, we also compare our results with previous LDA+U calculations\cite{Budai2014} as shown in the lower panel of Fig.~\ref{fig:naturecompare}. Comparison between the results obtained from SCAN and LDA+U clearly shows that the results obtained from SCAN serves better than the LDA+U. In particular, the peak features in the low frequency region around 20 meV is more intense compared to the LDA+U results as observed in experiments. 

We take a more closer look on the phonon density of states and study the effect of the V-V dimer on it. The role of V-V dimerization on the insulating behavior of the M1 phase is well understood now, whereas its effect on the lattice vibrations is still not extensively studied. Here, we aim to explore such physics. As we have already discussed that the V-V dimer distance deviates significantly from the experimental one, we compare phonon density of states of M$_1$ phase using both optimized and experimental structure as shown in Fig.~\ref{fig:ramancompare}. We observe that there are large number of negative frequencies in the low frequency region of the phonon density of states derived from experimental structure as displayed in upper panel of Fig.~\ref{fig:ramancompare}, whereas there are no such negative frequencies in the low frequency region of the phonon density of states considering optimized structure as shown in the lower panel of Fig.~\ref{fig:ramancompare}. The absence of such negative frequencies in the low frequency region of the phonon density of states corresponding to optimized structure is well expected. Moreover, there are also other noticeable differences. There is a peak in the low frequency region around 190 cm$^{-1}$ and a peak in the high frequency region around 609 cm $^{-1}$ in the phonon density of states obtained from with out optimizing the structure. The low frequency peak is mainly contributed from V-V pair and high frequency peak is originated from V-O pair which can be understood from Fig.~\ref{fig:partialdosM1}. The positions of the peak and nature of the vibrations are in excellent agreement with the observed Raman spectra. However, the position of the peaks are shifted in the phonon density of states corresponding to optimized structure compared to without optimization one. We attribute this to the drastic change in the short V-V dimer distance that occurred in the optimization process Table~\ref{tab:M1optimization}.      

Modulation of the V-V dimer length can be accomplished with strain engineering. It has previously been shown in VO$_2$/TiO$_2$ epitaxial thin films that tensile strain on the c$_R$-axis shifts the MIT mechanism towards bulk-like characteristics largely driven by the Peierls physics while elongation of the c$_R$-axis induces a larger influence from electron correlation or Mott physics.\cite{Quackenbush2016,Paez2020,Lee2019}. However, due to the strong signal from the TiO$_2$ in Raman spectroscopy, the strain effect on the phonon modes has been misinterpreted.\cite{Yang2016} Moreover, although strain has been accomplished on numerous substrates such as ZnO\cite{Wong2013}, Al$_2$O$_3$\cite{Fan2013}, LSAT\cite{Liu2018}, and RuO$_2$\cite{Fischer2020}, the formation of rotational domains may complicate fundamental structural analysis in the investigation of the MIT. Like TiO$_2$, MgF$_2$ is isostrucutral to VO$_2$ in its high temperature phase, and exhibits weaker Raman response than TiO$_2$ substrate\cite{doi:10.1063/1.4813442} and thus we selected it as the substrate for the epitaxial growth of VO$_2$ thin films. Fabrication of VO$_2$ on MgF$_2$ (001) induces the tensile strain of the c$_R$-axis while MgF$_2$ (110) films elongate the c$_R$-axis. Indeed, using Raman spectroscopy we observed spectral shift of the characteristic Raman peaks which represent V-V and V-O phonon modes. As shown in the right panel of Fig.~\ref{fig:lowramancompare} and Fig.~\ref{fig:highramancompare}, low frequency peak is shifted from 196 cm $^{-1}$ to 197 cm $^{-1}$, whereas high frequency peak is shifted from 609 cm ${-1}$ to 618 cm $^{-1}$ with the application of strain. Such spectral 
displacement is shown in our calculated phonon density of states by manipulating V-V dimmer.
As presented in the left panel of Fig.~\ref{fig:lowramancompare} and Fig.~\ref{fig:highramancompare}, the low frequency peak has been shifted from 190 $cm^{-1}$ to 230 cm$^{-1}$ as short dimer V-V distance is varied from 2.62 $\AA$ to 2.47 $\AA$. Similarly, the high frequency peak is deviated from 609 cm $^{-1}$ to 628 cm $^{-1}$. Thus, our calculations qualitatively explain the effect of strain on the lattice vibrations as observed in our Raman experiments.

\section{Conclusions}
In this paper, we have computed vibrational properties of the R and M$_1$ phases of VO$_2$ using first-principles calculations. A crucial step for calculating such vibrational properties is the structure optimization which is found to be non trivial for the M$_1$-VO$_2$. We have employed SCAN meta-GGA functional and successfully optimized the M$_1$ structure. The performance of the SCAN has been compared with various other functional in optimizing the M$_1$ structure. Particularly, the SCAN has been found to be as accurate as the Hybrid functional. The SCAN optimized structure has been shown to produce consistent electronic properties including both the R and M$_1$ phases of VO$_2$. We have shown that the R phase is a metal, whereas the M$_1$ phase is found to be an insulator with finite band gap value. Moreover, we numerically demonstrate that a gap can be opened up as the V-V dimer length is varied without incorporating electronic correlations. Hence, our results strongly suggest that the M$_1$ phase is not a conventional Mott insulator given the sensitivity to the dimer distance to its electronic phase. Thus, we have validated the SCAN for phonon calculations of VO$_2$. The phonon softening related with the R phase and the phonon stiffening of the M$_1$ phase explain the structural phase transition from the R to M$_1$ as temperature is lowered. We argue that such structural phase transition can play a vital role in the MIT of VO$_2$ as found in recent experiment. The V-V dimer has significant impact on both the electronic and vibrational properties of the M$_1$ phase. We show that the vibrational density of states significantly changes with the variation of the V-V dimer distance. Our calculated vibrational spectra are corroborated by previous neutron scattering experiment as well as our Raman experiment. Moreover, our first principles based calculations of bulk VO$_2$ enable us to gain insight into the structural properties of strained VO$_2$ as found in our Raman experiment. Our work also suggests that the SCAN can be utilized for investigating phonon properties of related materials and such work is in progress.    
        
\begin{acknowledgments} 
We thank Dr. Christopher N Singh for helpful discussions in our theoretical calculations, Dr. David J. Gosztola and  Dr. Benjamin T. Diroll for assistance with Raman microscopy measurements.  This paper is based on the work supported by the Air Force Office of Scientific Research under Award No. FA9550-18-1-0024 administered by Dr. Ali Sayir. For the film synthesis we acknowledge the National Science Foundation (Platform for the Accelerated Realization, Analysis, and Discovery of Interface Materials (PARADIM)) under Cooperative Agreement No. DMR-1539918. This research used resources of the Center for Nanoscale Materials, an Office of Science user facility, was supported by the U.S. Department of Energy, Office of Science, Office of Basic Energy Sciences, under Contract No. DE-AC02-06CH11357. Moreover, this work made use of the Cornell Center for Materials Research Shared Facilities which are supported through the NSF MRSEC program (DMR-1719875).
Galo J. Paez acknowledges doctoral degree grant support (Grant No. E0565514) from the Comisión Fulbright Ecuador in conjunction with the Ecuadorian national science department Secretaría de Educación Superior, Ciencia, Tecnología e Innovación (Senescyt).
\end{acknowledgments}

\bibliographystyle{apsrev4-1}
\bibliography{main}

\end{document}